\documentclass[
twocolumn,
prl,showpacs,
tightenlines,
nobibnotes,aps,
secnumarabic,
amsmath,amssymb,floatfix]{revtex4}
\usepackage{docs}
\renewcommand{\subsection}[1]{~\newline~{\bf{#1:}}}
\pacs{87.15-v,87.15.Aa,83.10,83.50.Ax,83.80.Rs,05.10-a,47.50.+d}

\newcommand{\dgamma}{\dot{\mbold{\gamma}}}
\newcommand{\bgam}{{\dot{\gamma}}}

\newcommand{\bcP}{\mbold{\cal P}}

\newcommand{\bde}{\begin{description}}

\newcommand{\ben}{\begin{enumerate}}
\newcommand{\beq}{\begin{eqnarray}}

\newcommand{\beqn}{\begin{eqnarray*}}

\newcommand{\bF}{{\bf F}}

\newcommand{\mbold}[1]{\mbox{\boldmath $#1$}}

\newcommand{\bqu}{\begin{quote}}
\newcommand{\br}{{\bf r}}

\newcommand{\bvb}{\begin{verbatim}}

\newcommand{\bxi}{\mbold{\xi}}

\newcommand{\ea}{\end{eqnarray}}

\newcommand{\ede}{\end{description}}
\newcommand{\een}{\end{enumerate}}
\newcommand{\eeq}{\end{eqnarray}}
\newcommand{\eeqn}{\end{eqnarray*}}

\newcommand{\equ}{\end{quote}}

\newcommand{\evb}{\end{verbatim}}

\newcommand{\<}{\langle}

\newcommand{\no}{{\bf{\hat n}}}

\newcommand{\oh }{\frac{1}{2}}

\newcommand{\para}{\parallel}

\newcommand{\suppress}[1]{}

\newcommand{\ta}{{\bf{\hat{t}}}}

\newcommand{\W}{{\cal W}}

\newcommand{\xhat}{\hat{\bf x}}

\newcommand{\yhat}{\hat{\bf y}}

\newcommand{\pmbeg}{\begin{pmatrix}}
\newcommand{\pmend}{\end{pmatrix}}

\renewcommand{\O}{{\cal {O}}}

\renewcommand{\>}{\rangle}

 \newif\ifpdf \ifx\pdfoutput\undefined
\pdffalse 
\else
\pdfoutput=1 
\pdftrue \fi

\ifpdf \usepackage[pdftex]{graphicx} \DeclareGraphicsExtensions{.pdf, .jpg, .tif}
\else \usepackage{graphicx}\DeclareGraphicsExtensions{.eps, .jpg}
\fi

\def\kB{k_B}

\def\rot{\mathsf{rot}}
\def\bstiff{\kappa}  
\def\bmode{\vartheta}  

\date{\today}
\begin{document}
\title{ The Stochastic Spectral Dynamics of Bending and Tumbling}
\author{Chris H. Wiggins,$^{1,2,5}$ Alberto Montesi,$^{3,4,5}$ Matteo Pasquali$^{3,4,5}$}
\affiliation{
$^1$Department of Applied Physics and Applied Mathematics and\\
$^2$Center for Computational Biology and Bioinformatics\\
Columbia University, New York NY 10027;\\
$^3$Department of Chemical Engineering and\\
$^4$Center for Nanoscale Science and Technology\\
Rice University, Houston TX 77005;\\
$^5$The Kavli Institute for Theoretical Physics, Santa Barbara CA 93106
}
\begin{abstract}
  Traditional models of wormlike chains in shear flows at finite
  temperature approximate the equation of motion via finite
  difference 
  discretization (bead and rod models).  We introduce here a
  new method 
  based on a spectral
  representation in terms of the natural eigenfunctions.
  This formulation separates tumbling and bending dynamics, clearly
  showing their interrelation, naturally orders the bending dynamics
  according to the characteristic decay rate of its modes, and
  displays coupling among bending modes in a general flow.
  This hierarchy naturally yields a low dimensional
  stochastic dynamical system which recovers and extends previous
  numerical results and which leads to a fast and efficient numerical
  method for studying the stochastic nonlinear dynamics of
  semiflexible polymers in general flows. 
  This formulation will be useful for studying other physical systems
  described by constrained stochastic partial differential equations.

\end{abstract}

\maketitle


In fields such as complex fluids and single-molecule biophysics, one
is often interested in dynamic and statistical properties of
conformations of {\it semiflexible biopolymers}, those for which
resistance to bending is comparable to or larger than thermal forcing.
These polymers may be considered as ``wrinkled'' rods, in contrast
with flexible polymers whose coiled configurations are modeled
as diffusive random walks in space \cite{DnE86}.
Examples of the two extremes are microtubules, responsible (among other roles)
for separating chromosomes during cell division, and long strands of DNA,
e.g., those used in the flow experiments of Chu and coworkers
\cite{Smith99}

Numerical models of semiflexible polymers often rely on the
bead-spring
or bead-rod \cite{Everaers99,Pasquali01,Pasquali02}
discretizations of the wormlike chain model.
Such discretization schemes were originally developed for modeling fully flexible polymers
as hundreds of statistically independent extensible
or inextensible
links
\cite{Fixman78,Liu89,GrassiaHinch96,OttingerBook}.
Modeling becomes progressively more complicated as the segments
are treated as rods, because the inextensibility constraint
must be enforced by tensions which are determined by an algebraic
(involving no time derivatives) constraint which couples the motion
of all the parts of the chain.  Introducing inextensibility presents
two additional challenges: multiplicative coupling of the
tension and the configuration of the polymer
\cite{GoldsteinLanger95, GoldsteinPowersWiggins98}, which
leads to {\it multiplicative} noise in the polymer equation
of motion \cite{OttingerBook}; and a subtle correlation between
the configuration of the links which comes from projecting out
the momenta degrees of freedom of a system which is constrained
to live on a non-Euclidian subspace \cite{Fixman74,Fixman78,Hinch94,Morse02}.
Both challenges can be met by using appropriate simulation algorithms
\cite{OttingerBook}---e.g., a mid-step algorithm \cite{Fixman78,Hinch94,GrassiaHinch96},
or an appropriate predictor-corrector algorithm \cite{Liu89}---and by introducing
appropriate ``metric'' forces into the equations of motion
\cite{Fixman74,Fixman78,Morse02} and computing them with efficient
procedures \cite{Pasquali02}.
Introducing a bending energy in the model yields further complications because
inextensibility couples the longitudinal and transversal chain dynamics.
The relaxation time of a bending mode of wavenumber $j$ is
$\tau_j \equiv \zeta_\perp L^4/(k_j^4 \bstiff)$ \cite{Shankar02}, whereas the
longest relaxation time of a rodlike chain is
$\tau_\rot \equiv \zeta_\perp L^3/(72 \kB T)$.  Here
$\zeta_\perp \approx 4 \pi \mu / \ln(L/\rho)$ is the perpendicular
friction coefficient, $L$ is the chain length, $\rho$ is the chain diameter,
$\mu$ is the liquid viscosity, $k_j \approx (2j+1)\pi$ is
the dimensionless relaxation rate of the $j$-th bending eigenmode,
$\bstiff$ is the chain bending stiffness, $\kB$ is Boltzmann's constant,
and $T$ is temperature.  In simulations, the fast shape fluctuations
of the chain must be resolved to sample correctly the Boltzmann distribution;
therefore, the simulation timestep must be roughly
$\Delta_t \approx 0.1 \zeta_\perp L^4/((2N+1)^4 \pi^4 \bstiff)$, where $N$ is the shortest
wavenumber that can be captured by the discretized model and coincides
with the number of rods (minus one) in a bead-rod model.  Simulating the terminal relaxation
of a chain while resolving the fast modes requires $\approx \tau_\rot / \Delta_t$
timesteps, i.e, $\approx 10 (2N+1)^4 \pi^4 L_p / (72 L)$, where
$L_p \equiv \bstiff / \kB T$ is the persistence length of the chain
\cite{note1}.  For the calculation of linear viscoelastic properties, such a hurdle
was overcome by simulating the short-time behavior of semiflexible molecules with
finely discretized chains, and collapsing the results with those obtained by
simulating the long-time behavior with coarsely-discretized chains
\cite{Everaers99,Pasquali01}.  Such strategy, however, does not work
when applied to nonlinear viscoelastic properties and general flows and
deformations, because semiflexible chains buckle when subjected to
strong velocity gradients \cite{Alberto}; thus, in nonlinear simulations
one has to retain the fine discretization (in order to resolve curvature
during buckling) while also simulating long timescales (in order to capture
the steady rheological behavior).  Therefore, a new, effective method is needed
for studying flows of stiff chains at finite temperature.

As in earlier studies, the starting point is the equation of motion
for a continuous curve of position $\br(s,t)$ as a function of
arclength $s$ and time $t$, of bending stiffness
$\bstiff$, with inextensibility enforced by a tension $\Lambda$, moving in
a viscous liquid
at temperature $T$ and velocity
${\bf  U}\approx {\br\cdot\nabla U}=\dgamma\br$
\cite{KnR,Batchelor70}:
\beq
\partial_t\br-\dgamma\br&=& \bcP\left(\partial_s\bF_{\rm
    elastic}
  +\partial_s\bF_{\rm tension}+\bxi\right)
\label{eom}
\eeq
where $\<\bxi(s,t)\bxi(s',t')\> = 2 \kB T \bcP^{-1}\delta(t-t')\delta(s-s')$,
$\bF_{\rm elastic}=-\bstiff\partial_s^3\br$,
$\bF_{\rm tension}=\Lambda\partial_s\br$,
and the tension $\Lambda$ is determined by satisfying the inextensibility
constraint $|\partial_s \br| = 1$.
The mobility
tensor $\bcP\equiv c_\perp \no
\no+c_\para\ta\ta$ enforces the anisotropy associated with slender
bodies in Stokes flow ($c_\perp \equiv 1/\zeta_\perp$ and
$c_\para \equiv 2 c_\perp$ for a slender cylinder).
Unlike in previous studies, we work with the Eq.~\ref{eom} rather
than discretizing it into a bead-rod formulation.
We pose the dynamic in two dimensions, separating
bending from tumbling without constraining the position or orientation
of the rod whose configuration is the base state of our perturbation.
Differentiating Eq.~\ref{eom} and recasting the results in
terms of tangent and normal vectors
$\ta \equiv \partial_s\br=(\cos\vartheta,\sin\vartheta)$,
$\no \equiv (-\sin\vartheta,\cos\vartheta)$,
yields
\beq
\vartheta_t\no-\dgamma\ta = 
    \partial_s\bcP\left(\partial_s\bF_{\rm
    elastic}+\partial_s\bF_{\rm tension}+\bxi\right).  \nonumber
\eeq
Function\-a\-lly
differenti\-ating the elastic energy $E=(\bstiff/2)\int_0^L
ds(\partial_s\vartheta)^2$
yields the e\-lastic force;
whereas the orientation $\vartheta$ does not appear in the energy,
the curvature $\partial_s\vartheta$ is suppressed; this introduces
a natural perturbation parameter
$\varepsilon^2\equiv \kB T L / \bstiff \equiv L/L_p$.
Neglecting terms nonlinear in the curvature $\vartheta_s$
or its derivatives,
\beq \vartheta_t\approx
\no\dgamma\ta(\vartheta)
-
c_\perp \bstiff
\partial_s^4\vartheta+\ldots
\eeq
suggests representing
$\vartheta$ as a combination of eigenfunctions of the leading order operator:
$L^4\partial_s^4\W^\alpha=k_\alpha^4\W^\alpha$.  These
biharmonic eigenfunctions generalize many properties of Fourier
modes and, for high mode number, asymptote to the Fourier basis
\cite{LnLv7,Shankar02,GoldsteinPowersWiggins98}.
However, they satisfy the boundary conditions (no traction and
no torque) imposed on the ends of an elastic rod immersed in a
viscous fluid: $\partial_s
\W^\alpha=\partial_s^2\W^\alpha=0$.
The only allowed $0-$mode is independent of $s$. In this way, the
biharmonic eigenfunction basis naturally yields the
representation
\beq
\nonumber
\vartheta (t,s) \equiv \vartheta_0(t)\W^0(s)+\bmode_\alpha(t)\W^\alpha(s)\equiv
\vartheta_0
+\bmode_\alpha
\W^\alpha
\eeq
where $\vartheta_0(t)$ is the rod or tumbling term
and the $\bmode_\alpha(t)$ are a dynamically ordered sequence
of bending terms (summation implied).
Adjoint to the set of shape functions is a set of weight
eigenfunctions $\W_\alpha$ of conjugate boundary conditions
$\W_\alpha=\partial_s^3 \W_\alpha=0$. The
$\{\W^\alpha,\W_\alpha\}$ are biorthogonal
$\int ds \W_\alpha \W^\beta=L\delta_\alpha^\beta,$
a crucial observation for 
equipartition of energy,
\beq \frac{E}{k_BT}=
\oh \varepsilon^{-2} \sum k_\alpha^2\bmode_\alpha^2
\label{Equipartition}
\eeq
which yields immediately the equilibrium variance 
of any mode $\<\bmode_\alpha^2\>=
(\varepsilon/k_\alpha)^2$. This representation offers a natural
way of capturing the long wavelength bending modes with
accuracy while suppressing the short-wavelength modes (which
are uninportant in hydrodynamics) by truncating at a small $\alpha$.
The adjoint functions form a natural basis for the tension, which
obeys $\Lambda=0$ at edges:
$\Lambda \equiv \Lambda^\alpha \W_\alpha$

\begin{figure}[tbp]
\centerline{\includegraphics[width=3.25in] {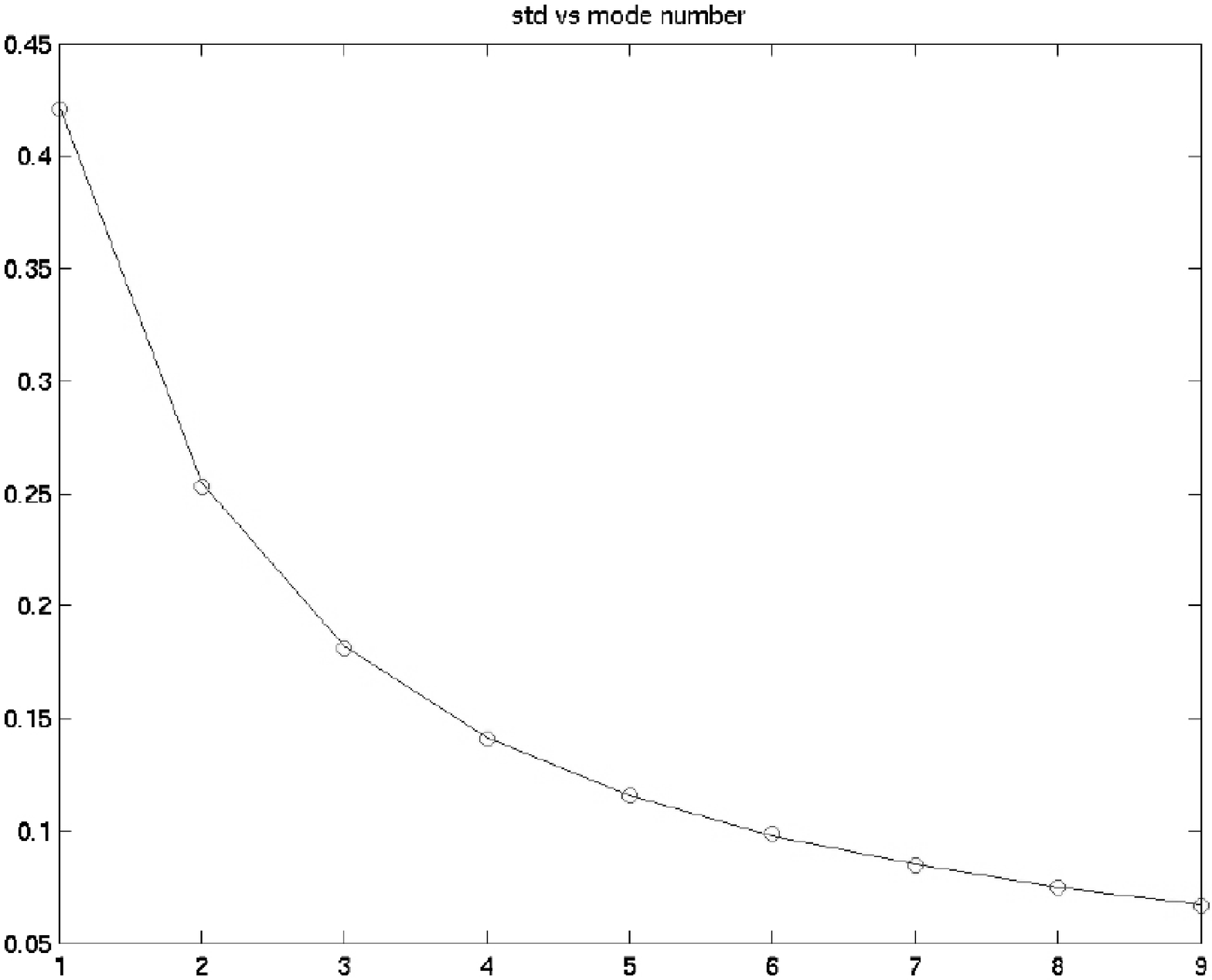}}
\caption{Theoretical curve ($\sqrt{\kB TL/(\kappa k^2)}$, Eq.~\ref{Equipartition})
and computed values of $\sqrt{\<\bmode_\alpha^2\>}$, first 9
bending modes.}
\label{fig:boltzmann}
\end{figure}

The dynamics of the modes and the constraint can be derived
in simple shear ($\dgamma=\bgam \xhat\yhat^T$)
by the Galerkin method by multiplying Eq.~\ref{eom} by the appropriate
weight functions ($\W_\alpha$ for the modes, $\W^\alpha$ for the tensions),
integrating along the chain contour, taking the limit for small $\varepsilon$,
and dropping terms
$\O(\epsilon^2)$ from the $\Lambda$ equation and $\O(\epsilon^3)$ from the
$\dot{\bmode}$ equation,
\beq
\label{eommode}
0&=&-c_\para k_\beta^2\Lambda^\beta+ [
\ta\dgamma\no]
_\beta+[\eta^\para]_\beta
\\
\nonumber
\dot{\vartheta}_0&=&-\bgam\left(\sin^2\vartheta_0+\cos(2\vartheta_0)[\bmode^2]_0\right)+\bmode_\alpha\Lambda^\gamma\Xi^\alpha_{0\gamma}+[\eta^\perp]_0\\
\nonumber
\dot{\bmode}_\beta&=&[
\no\dgamma\no ]_\beta
-\frac{c_\perp \bstiff}{L^4}
k_\beta^4\bmode_\beta+\bmode_\alpha\Lambda^\gamma
\Xi^\alpha_{\beta\gamma}+[\eta_\perp]_\beta
\eeq
where summation over $\alpha,\gamma$ is implied,
$\eta^{\{\para,\perp\}}= \{\ta,\no\}\cdot\partial_s\bxi$,
projection onto the $\beta$ mode is indicated by $[\cdots]_\beta$,
and $\Xi$ is an overlap integral of the basis functions which depends
on the mobility constants $\{c_\para,c_\perp\}$. For isotropic drag,
for example, $\Xi^\alpha_{\beta\gamma}\propto \int ds
\W^\gamma\left(k_\beta^2\W^\beta\W_\alpha-k_\alpha^2\W^\alpha\W_\beta\right)$.

   { \it{Stability --}}
    Shear-induced buckling, in which shear and elasticity compete to
    determine the stability of a non-Brownian rod at fixed $\vartheta_0$,
    has been studied analytically, experimentally
    \cite{FM}, and numerically
    \cite{BeckerShelley01}.
    The numerical results
    show that in simple shear the
    instability occurs when
    $
    (\bgam_* L^4\sin 2\vartheta_0)/(c_\para \bstiff)=-153.2$.
    We derive analytically this criterion by expanding the shear terms
\beq
    \gamma_\perp(\vartheta)\equiv\no\dgamma\no\approx
    \gamma_\perp(\vartheta_0)+\bmode_\alpha
\W^\alpha\gamma_\perp'(\vartheta_0) 
\eeq
($+\O(\bmode^2)$)---in simple shear
$\gamma_\perp(\vartheta_0)\propto-\sin^2\vartheta_0$.
Expanding the equation of motion and dropping $\bmode$- and $T-$dependent terms from the
tension, the linear instability criterion is
\beq
0=\left(\bgam_*\sin 2\theta_0- \frac{c_\perp \bstiff}{L^4} k_\beta^4
\right)\bmode_\beta+\frac{\bgam_*}{6}\Xi^\alpha_{\beta0}
\bmode_\alpha
\eeq
The overlap integral $\Xi^\alpha_{\beta0}$ is
well-approximated by its diagonal contribution $\Xi^\beta_{\beta0}\delta_\beta^\alpha$.
Using the large-$\alpha$ Fourier approximation of the
basis functions yields the instability criterion $ \bgam_* L^4\sin
2\vartheta_0/(c_\para \bstiff)= -162\pi^4/(76+3\pi^2)=-149.4$, with an
error of $2.6\%$. Calculating the diagonal overlap integral reduces
the error to $.2\%$ \cite{note2}.

\begin{figure}[tbp]
\centerline{\includegraphics[width=3.00in] {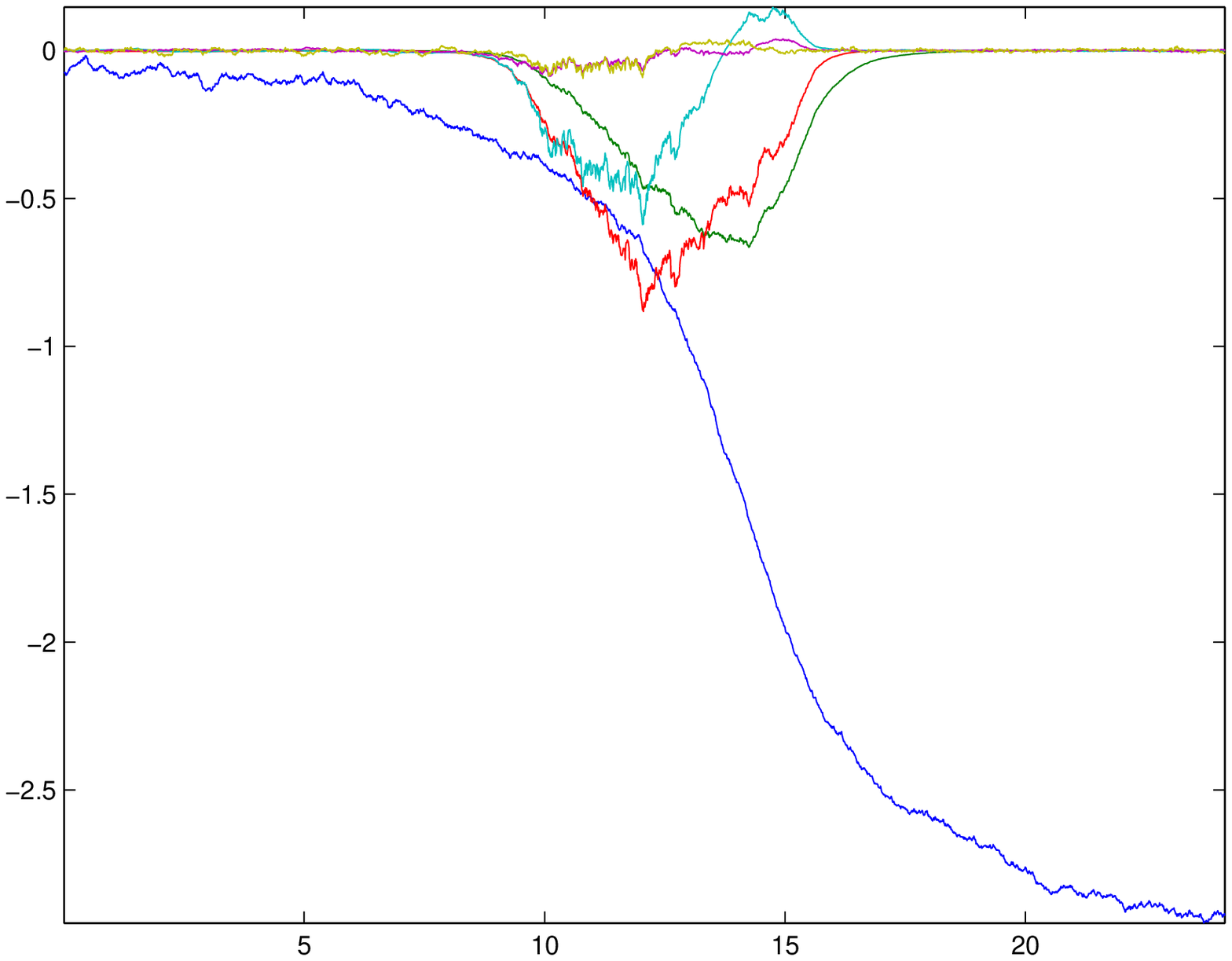}}
\caption{Stochastic spectral dynamics of bending and tumbling. The curve which leaves the
origin traces $\vartheta_0(t)$.
The remaining curves describe the bending modes
$\bmode_\alpha(t)$. Each successive mode is smaller in
autocorrelation, as expected from the equipartion (Eq.~\ref{Equipartition}).
Here $\bgam=8.5\cdot 10^3 \kappa L^4/(c_\perp)$, $L=L_p$,  and
the x-axis is $t\bgam$.}
\label{fig:dynsys}
\end{figure}

\begin{figure}[tbp]
\centerline{\includegraphics[width=3.00in] {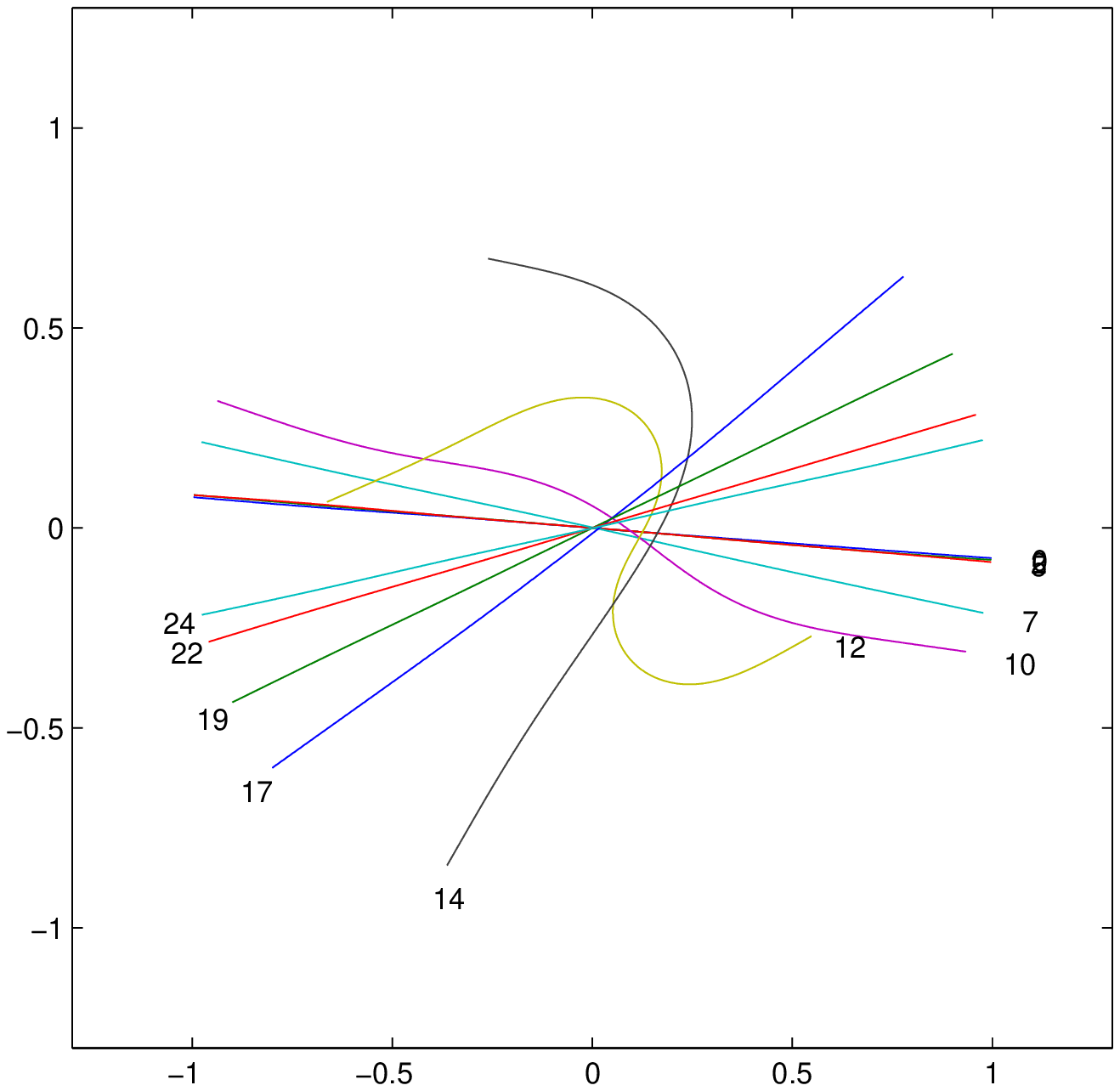}}
\caption{The same dynamics, represented in real space and imaged at several representative
values of $t \bgam$. The complex
motion is captured by only 6 
degrees of freedom.}
\label{fig:tumble}
\end{figure}

    {\it{Finite temperature --}}
    In shear flow, the zero-temperature dynamics asymptote to a straight rod
    aligned in the velocity direction, because the equation of change
    of $\vartheta_0$ reduces to
    ${\dot \vartheta}_0 = -\bgam\sin^2\vartheta_0 \rightarrow 0$
    as $\vartheta_0 \rightarrow j \pi$.
    In the presence of stochastic forces, the dynamics is richer
    because the $\vartheta_0= 0$ solution is nonlinearly unstable;
    thus, the stochastic forces episodically drive $\vartheta_0$ to
    initiate tumbling events, assisted by shear-induced
    buckling en route.
To study the
    fully-nonlinear stochastic dynamics we use a midstep algorithm,
    consistent with variable diffusivity and Stratonovich-interpreted noise
    \cite{KlodenPlaten94,note3}.
    In the absence of shear, the algorithm yields the correct Boltzmann
    distribution of bending modes (Fig.~\ref{fig:boltzmann}).
    Illustrative dynamics in the presence of shear are shown in
    Figs.~\ref{fig:dynsys} and~\ref{fig:tumble}.


{\it{Extensions and connections --}}
One motivation for this work is better understanding of
complex fluids.  In dilute suspensions, the
contribution of molecules to the stress tensor can be calculated
via the continuous generalization for wormlike chains \cite{Morse98a}
of the Kirkwood formula \cite{Kirkwood67}.
Quantitative comparison between the efficient spectral numerics and
calculations of the stress and molecular conformation
(e.g., gyration tensor) by established but inefficient large-scale simulations
of bead-rod chains is in progress \cite{Alberto}.
Albeit restricted to two dimensions, the fully nonlinear stochastic method
is completely general with respect to time-dependent flows and kinematics
(shear, extensional, mixed); thus, it will be useful for comparison
with
``single-bead rheology'' studies in actin gels \cite{GittesMacKintosh98}.
An extension of the spectral method to include the full three-dimensional
representation of semiflexible chains in general flows is underway.

An additional
topic for future work is the
application
of stochastic mode-elimination
\cite{kurtz,papanicolaou,eve1,eve2},
integrating out the 
fastest bending modes.  Mode elimination
renormalizes the noise statistics, and can be used to
yield a low-dimensional partial differential
equation for the probability density.  Because the properties
of interest for biopolymers (e.g., configurational changes
connected with macroscopic rheology) evolve on slow timescales,
stochastic mode elimination is an excellent
candidate for mathematically sound yet experimentally relevant
analysis.  Such a foray would be quite daunting without simple, low
dimensional stochastic dynamical systems as in Eq.~\ref{eommode} for which
the mapping from Langevin descriptions of trajectories to
Fokker-Planck descriptions of probabilities is well-developed.

{\it{Acknowledgments --}}
We thank David Morse, Jonathan Mattingly, and Guillaume Bal for useful discussions.
Partial financial support was provided by NSF under grant CTS-CAREER-0134389.

\bibliography{biopoly}

\begin{thebibliography}{33}
\bibitem[{not({\natexlab{c}})}]{note3}
\bibinfo{note}{We remove stiffness as in \cite{HouLowengrubShelley94} and also
  speed-up the numerics by Taylor-expanding the shear terms and thus storing
  the overlap integrals (otherwise integrals such as $\int ds \W_\beta
  \sin^2\vartheta(s)$ must be recomputed at each iteration).}
\end{thebibliography}




\end{document}
\end